\documentclass[amsmath,aps,showpacs,a4paper,10pt]{revtex4}

 \usepackage{epsf}
 \usepackage{graphicx}    

\begin{document}

 \newcommand{\be}[1]{\begin{equation}\label{#1}}
 \newcommand{\ee}{\end{equation}}
 \newcommand{\bea}{\begin{eqnarray}}
 \newcommand{\eea}{\end{eqnarray}}
 \def\disp{\displaystyle}

 \def\gsim{ \lower .75ex \hbox{$\sim$} \llap{\raise .27ex \hbox{$>$}} }
 \def\lsim{ \lower .75ex \hbox{$\sim$} \llap{\raise .27ex \hbox{$<$}} }

 \begin{titlepage}

 \begin{flushright}
 arXiv:0708.1894
 \end{flushright}

 \title{\Large \bf Cosmological Constraints on
 New Agegraphic Dark Energy}

 \author{Hao~Wei\,}
 \email[\,email address:\ ]{haowei@mail.tsinghua.edu.cn}
 \affiliation{Department of Physics and Tsinghua Center for
 Astrophysics,\\ Tsinghua University, Beijing 100084, China}

 \author{Rong-Gen~Cai\,}
 \email[\,email address:\ ]{cairg@itp.ac.cn}
 \affiliation{Institute of Theoretical Physics, Chinese Academy of
 Sciences, P.O. Box 2735, Beijing 100080, China}

 \begin{abstract}\vspace{1cm}
 \centerline{\bf ABSTRACT}\vspace{2mm}
In this work, we consider the cosmological constraints on the
 new agegraphic dark energy (NADE) proposed in arXiv:0708.0884,
 by using the observational data of type~Ia supernovae (SNIa),
 the shift parameter from cosmic microwave background (CMB) and
 the baryon acoustic oscillation (BAO) peak from large scale
 structures (LSS). Thanks to its special analytic features
 in the radiation-dominated and matter-dominated epochs, NADE
 is a {\em single-parameter} model in practice because once
 the single model parameter $n$ is given, all other physical
 quantities of NADE can be determined correspondingly. The
 joint analysis gives the best-fit value (with $1\sigma$
 uncertainty) $n=2.716^{+0.111}_{-0.109}$, and the derived
 $\Omega_{m0}$, $\Omega_{q0}$ and $w_{q0}$ (with $1\sigma$
 uncertainties) are $0.295^{+0.020}_{-0.020}$,
 $0.705^{+0.020}_{-0.020}$ and $-0.794^{+0.006}_{-0.005}$,
 respectively. In addition, we find that the coincidence
 problem could be solved naturally in the NADE model provided
 that $n$ is of order unity.
 \end{abstract}


 \maketitle

 \end{titlepage}

 \renewcommand{\baselinestretch}{1.6}



\section{Introduction}\label{sec1}
In general relativity, one can measure the spacetime without any
 limit of accuracy.  In quantum mechanics, however, the well-known
 Heisenberg uncertainty relation puts a limit of accuracy in these
 measurements. Following the line of quantum fluctuations of
 spacetime, K\'{a}rolyh\'{a}zy and his collaborators~\cite{r1}
 (see also~\cite{r2}) made an interesting observation concerning
 the distance measurement for Minkowski spacetime through a
 light-clock {\it Gedanken experiment}, namely, the distance $t$
 in Minkowski spacetime cannot be known to a better accuracy than
 \be{eq1}
 \delta t=\lambda t_p^{2/3}t^{1/3},
 \ee
 where $\lambda$ is a dimensionless constant of order
 unity~\cite{r2,r3}. We use the units $\hbar=c=k_B=1$
 throughout this work. Thus, one can use the terms like length
 and time interchangeably, whereas $l_p=t_p=1/m_p$ with $l_p$,
 $t_p$ and $m_p$ being the reduced Planck length, time and
 mass, respectively.

The K\'{a}rolyh\'{a}zy relation~(\ref{eq1}) together with the
 time-energy uncertainty relation enables one to estimate an
 energy density of the metric quantum fluctuations of Minkowski
 spacetime~\cite{r3,r2}. Following~\cite{r3,r2}, with respect to
 Eq.~(\ref{eq1}) a length scale $t$ can be known with a maximum
 precision $\delta t$ determining thereby a minimal detectable cell
 $\delta t^3\sim t_p^2 t$ over a spatial region $t^3$. Such a cell
 represents a minimal detectable unit of spacetime over a given
 length scale $t$. If the age of the Minkowski spacetime is $t$,
 then over a spatial region with linear size $t$ (determining the
 maximal observable patch) there exists a minimal cell $\delta t^3$
 the energy of which due to time-energy uncertainty relation can not
 be smaller than~\cite{r3,r2}
 \be{eq2}
 E_{\delta t^3}\sim t^{-1}.
 \ee
 Therefore, the energy density of metric fluctuations of
 Minkowski spacetime is given by~\cite{r3,r2}
 \be{eq3}
 \rho_q\sim\frac{E_{\delta t^3}}{\delta t^3}\sim
 \frac{1}{t_p^2 t^2}\sim\frac{m_p^2}{t^2}.
 \ee
 We refer to the original papers~\cite{r3,r2} for more details.
 It is worth noting that in fact, the K\'{a}rolyh\'{a}zy
 relation~(\ref{eq1}) and the corresponding energy
 density~(\ref{eq3}) have been independently rediscovered later
 for many times in the literature (see e.g.~\cite{r4,r5,r6}).

Based on the energy density~(\ref{eq3}), a so-called agegraphic dark
 energy~(ADE) model was proposed in~\cite{r7}. There, as the most
 natural choice, the time scale $t$ in Eq.~(\ref{eq3}) is chosen to
 be the age of the universe
 \be{eq4}
 T=\int_0^a\frac{da}{Ha},
 \ee
 where $a$ is the scale factor of our universe; $H\equiv\dot{a}/a$
 is the Hubble parameter; an over dot denotes the derivative with
 respect to cosmic time. Then, the ADE model was extended to include
 the interaction between ADE and background matter~\cite{r8}. The
 statefinder diagnostic and $w-w^\prime$ analysis for the ADE models
 without and with interaction were performed in~\cite{r9}. The ADE
 was constrained by using some old high redshift objects~\cite{r10}
 and type Ia supernovae~(SNIa), the shift parameter from cosmic
 microwave background~(CMB) and the baryon acoustic oscillation~(BAO)
 peak from large scale structures~(LSS)~\cite{r11}.

On the other hand, a new model of ADE was proposed in~\cite{r12},
 where the time scale in Eq.~(\ref{eq3}) is chosen to be the
 conformal time $\eta$ instead of the age of the universe. This
 new agegraphic dark energy~(NADE) has some new features different
 from the ADE proposed in~\cite{r7}. We will briefly review the main
 points of NADE model in the next section.

Thanks to its special analytic features in the radiation-dominated
 and matter-dominated epochs, NADE is a {\em single-parameter} model
 in practice, unlike the {\em two-parameters} ADE model~\cite{r11}
 (this point will be explained below). If the single model parameter
 $n$ is given, all other physical quantities of NADE are determined
 correspondingly. In this work, we find that the coincidence problem
 could be solved naturally in the NADE model provided that $n$ is
 of order unity. In addition, we constrain the NADE by using the
 cosmological observations of SNIa, CMB and LSS. Different from the
 cosmological constraints on ADE~\cite{r11}, the single parameter
 $n$ of NADE model can be strictly constrained. Then, the other
 physical quantities of NADE can also be determined tightly.


\section{The model of new agegraphic dark energy}\label{sec2}

\subsection{Main points of the NADE model}\label{sec2a}
In the NADE model~\cite{r12}, we choose the time scale in
 Eq.~(\ref{eq3}) to be the conformal time $\eta$ instead, which is
 defined by $dt=a\,d\eta$ [where $t$ is the cosmic time, do not
 confuse it with the $t$ in Eq.~(\ref{eq3})]. Therefore, the energy
 density of NADE reads
 \be{eq5}
 \rho_q=\frac{3n^2m_p^2}{\eta^2},
 \ee
 where the numerical factor $3n^2$ is introduced to parameterize
 some uncertainties, such as the species of quantum fields in
 the universe, the effect of curved spacetime [since the energy
 density in Eq.~(\ref{eq3}) is derived for Minkowski spacetime],
 and so on. Since both the numerical factor $\lambda$ in
 Eq.~(\ref{eq1}) and the coefficient in time-energy uncertainty
 relation Eq.~(\ref{eq2}) are of order unity, it is anticipated
 that the parameter $n$ in Eq.~(\ref{eq5}) which comes from
 Eqs.~(\ref{eq1}) and~(\ref{eq2}) is also of order unity. The
 conformal time $\eta$ is given by
 \be{eq6}
 \eta\equiv\int\frac{dt}{a}=\int\frac{da}{a^2H}.
 \ee
 If we write $\eta$ to be a definite integral, there will be an
 integration constant in addition. Note that $\dot{\eta}=1/a$.

In this work, we consider a flat Friedmann-Robertson-Walker~(FRW)
 universe containing NADE and pressureless matter. Note that the
 assumption of flatness is motivated by the inflation scenario
 and is also consistent with the observation of CMB~\cite{r26}.
 The corresponding Friedmann equation reads
 \be{eq7}
 H^2=\frac{1}{3m_p^2}\left(\rho_m+\rho_q\right),
 \ee
 where $\rho_m$ is the energy density of pressureless matter.
 It is convenient to introduce the fractional energy densities
 $\Omega_i\equiv\rho_i/(3m_p^2H^2)$ for $i=m$ and $q$. From
 Eq.~(\ref{eq5}), the corresponding fractional energy density
 of NADE is given by
 \be{eq8}
 \Omega_q=\frac{n^2}{H^2\eta^2}.
 \ee
 By using Eqs.~(\ref{eq5})---(\ref{eq8}) and the energy
 conservation equation $\dot{\rho}_m+3H\rho_m=0$, we find that
 the equation of motion for $\Omega_q$ is given by~\cite{r12}
 \be{eq9}
 \frac{d\Omega_q}{da}=\frac{\Omega_q}{a}\left(1-\Omega_q\right)
 \left(3-\frac{2}{n}\frac{\sqrt{\Omega_q}}{a}\right).
 \ee
 From the energy conservation equation
 $\dot{\rho}_q+3H(\rho_q+p_q)=0$, as well as Eqs.~(\ref{eq8})
 and~(\ref{eq5}), it is easy to find that the equation-of-state
 parameter~(EoS) of NADE $w_q\equiv p_q/\rho_q$ is given
 by~\cite{r12}
 \be{eq10}
 w_q=-1+\frac{2}{3n}\frac{\sqrt{\Omega_q}}{a}.
 \ee
 Obviously, the scale factor $a$ enters Eqs.~(\ref{eq9})
 and~(\ref{eq10}) explicitly. When $a\to\infty$, $\Omega_q\to 1$,
 thus $w_q\to -1$ in the late time. When $a\to 0$, $\Omega_q\to 0$,
 we cannot obtain $w_q$ from Eq.~(\ref{eq10}) directly. Let us
 consider the matter-dominated epoch,
 $H^2\propto\rho_m\propto a^{-3}$. Thus,
 $a^{1/2}da\propto dt=ad\eta$. Therefore, $\eta\propto a^{1/2}$.
 From Eq.~(\ref{eq5}), $\rho_q\propto a^{-1}$.  From the energy
 conservation equation $\dot{\rho}_q+3H\rho_q(1+w_q)=0$, we obtain
 that $w_q=-2/3$ in the matter-dominated epoch. Since
 $\rho_m\propto a^{-3}$, it is expected that $\Omega_q\propto a^2$.
 Comparing $w_q=-2/3$ with Eq.~(\ref{eq10}), we find that
 $\Omega_q=n^2a^2/4$ in the matter-dominated epoch as expected.
 For $a\ll 1$, if $n$ is of order unity, $\Omega_q\ll 1$ naturally.
 On the other hand, one can check that $\Omega_q=n^2a^2/4$ satisfies
 $$\frac{d\Omega_q}{da}=\frac{\Omega_q}{a}
 \left(3-\frac{2}{n}\frac{\sqrt{\Omega_q}}{a}\right),$$
 which is the approximation of Eq.~(\ref{eq9}) for
 $1-\Omega_q\simeq 1$. Therefore, all things are consistent.

\newpage 

In fact, we can further extend our discussion to include the
 radiation-dominated epoch~\cite{r12}. However, since our main
 aim of this work is to constrain NADE by using the
 cosmological observations of SNIa, CMB and LSS which only
 concern the matter-dominated epoch, here we skip the
 radiation-dominated epoch. Instead, we only give a brief
 summary of the results obtained in~\cite{r12}, namely, in the
 radiation-dominated epoch, $w_q=-1/3$ whereas $\Omega_q=n^2a^2$;
 in the matter-dominated epoch, $w_q=-2/3$ whereas
 $\Omega_q=n^2a^2/4$; eventually, the NADE dominates; in the
 late time $w_q\to -1$ when $a\to\infty$, the NADE mimics
 a cosmological constant. We refer to the original
 paper~\cite{r12} for more details.


 \begin{center}
 \begin{figure}[htbp]
 \centering
 \includegraphics[width=0.97\textwidth]{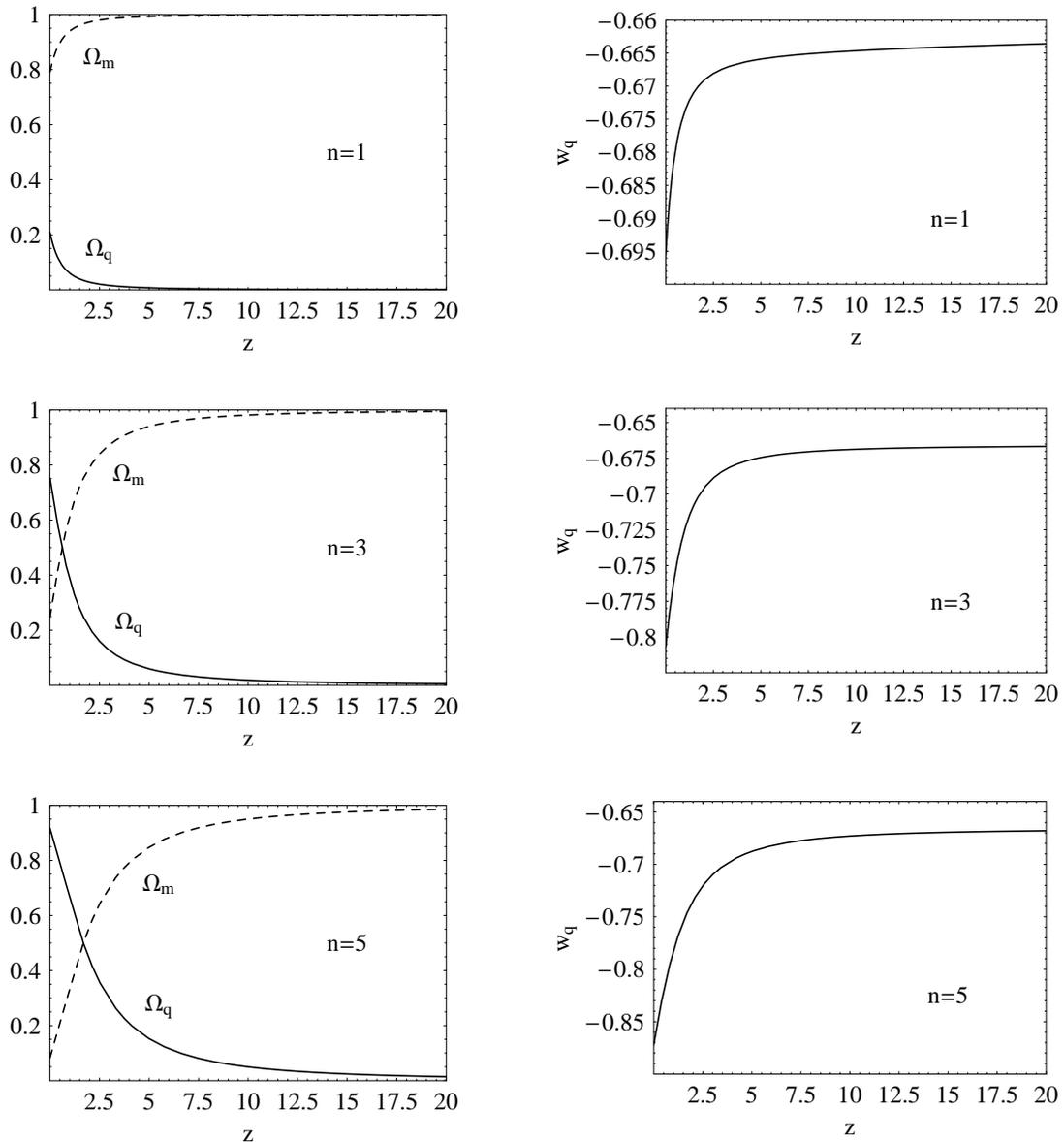}
 \caption{\label{fig1} The resulting $\Omega_q(z)$,
 $\Omega_m(z)=1-\Omega_q(z)$ and $w_q(z)$ obtained from
 Eqs.~(\ref{eq11}) and~(\ref{eq10}) with the initial
 condition $\Omega_q(z_{ini})=n^2(1+z_{ini})^{-2}/4$ at
 $z_{ini}=2000$ for different $n$.}
 \end{figure}
 \end{center}


\subsection{NADE and the coincidence problem}\label{sec2b}
It is worth noting that both $\Omega_q=n^2a^2$ and $n^2a^2/4 \ll 1$
 naturally in radiation-dominated and matter-dominated epochs where
 $a\ll 1$. These results are obtained without any additional
 assumption. It is natural to consider the coincidence problem. In
 many dark energy models, especially the $\Lambda$CDM model,
 $\Omega_{de}$ should be fine-tuned to be extremely small in the
 early time, in order to interpret the fact that the energy
 densities of dark energy and matter are comparable today. In some
 dynamical dark energy models, the tracker behavior or attractors
 in the interacting dark energy models are used to alleviate the
 coincidence problem.

Here, we see that the NADE model could shed new light on the
 coincidence problem in a {\em different} way. The key point is that
 as mentioned above $\Omega_q$ of NADE can be obtained analytically
 in both radiation-dominated and matter-dominated epochs. In
 particular, $\Omega_q=n^2a^2/4=n^2(1+z)^{-2}/4$ in the
 matter-dominated epoch, where $z=a^{-1}-1$ is the redshift (we set
 $a_0=1$ throughout; the subscript ``0'' indicates the present value
 of the corresponding quantity). Therefore, we can use
 $\Omega_q(z_{ini})=n^2(1+z_{ini})^{-2}/4$ at any $z_{ini}$ which is
 deep enough into the matter-dominated epoch to be the initial
 condition to solve the differential equation of $\Omega_q$.
 We rewrite Eq.~(\ref{eq9}) as
 \be{eq11}
 \frac{d\Omega_q}{dz}=-\Omega_q\left(1-\Omega_q\right)
 \left[3(1+z)^{-1}-\frac{2}{n}\sqrt{\Omega_q}\right].
 \ee
 As well-known, the energy densities of matter and radiation is
 equal at redshift $z_{eq}$ which is given by
 $1+z_{eq}\simeq 2.32\times 10^4\,\Omega_{m0}h^2$~\cite{r13}, where
 $h$ is defined by the Hubble constant $H_0=100\,h\,{\rm km/s/Mpc}$.
 For instance, $z_{eq}\simeq 3409$ for $\Omega_{m0}=0.3$ and
 $h=0.7$~\cite{r14}, whereas $z_{eq}\simeq 2505$ for
 $\Omega_{m0}=0.3$ and $h=0.6$~\cite{r15}. Therefore, we choose
 $z_{ini}=2000$ which is deep enough into the matter-dominated epoch.
 In fact, one can check that if we use initial condition
 $\Omega_q(z_{ini})=n^2(1+z_{ini})^{-2}/4$ at $z_{ini}=2000$,
 the resulting $\Omega_q(z)$ from Eq.~(\ref{eq11}) for $n=3$
 relatively deviates from $n^2(1+z)^{-2}/4$ less than approximately
 $1\%$ in the large interval $10\leq z\leq 2000$. Thus, if one uses
 other $z_{ini}$, the results deviate from the ones of $z_{ini}=2000$
 negligibly. This also justifies the choice of initial condition
 $\Omega_q(z_{ini})=n^2(1+z_{ini})^{-2}/4$ at $z_{ini}=2000$.
 In~Fig.~\ref{fig1}, we present the resulting $\Omega_q(z)$,
 $\Omega_m(z)=1-\Omega_q(z)$ and $w_q(z)$ from Eqs.~(\ref{eq11})
 and~(\ref{eq10}) for different $n$. We only plot the redshift range
 $0\leq z\leq 20$ in Fig.~\ref{fig1}, since $\Omega_q\simeq 0$ and
 $w_q\simeq-2/3$ closely in the range $20\leq z\leq 2000$. From
 Fig.~\ref{fig1}, it is easy to see that $\Omega_{q0}$ and
 $\Omega_{m0}=1-\Omega_{q0}$ are comparable provided that $n$ is of
 order unity. We stress that this result is obtained very naturally,
 without any additional assumption. Therefore, the coincidence
 problem could be solved naturally in the NADE model.


\section{Constraints on NADE from SNIa}\label{sec3}
In this section, we constrain NADE by using the latest 182 SNIa Gold
 dataset~\cite{r16}. The latest 182 SNIa Gold dataset compiled
 in~\cite{r16} provides the apparent magnitude $m(z)$ of the
 supernovae at peak brightness after implementing corrections for
 galactic extinction, K-correction, and light curve width-luminosity
 correction. The resulting apparent magnitude $m(z)$ is related to
 the luminosity distance $d_L(z)$ through (see e.g.~\cite{r17})
 \be{eq12}
 m_{th}(z)=\bar{M}(M,H_0)+5\log_{10} D_L(z),
 \ee
 where
 \be{eq13}
 D_L(z)=(1+z)\int_0^z \frac{d\tilde{z}}{E(\tilde{z};{\bf p})}
 \ee
 is the Hubble-free luminosity distance $H_0 d_L/c$ in a spatially
 flat FRW universe ($c$ is the speed of light); ${\bf p}$ denotes
 the model parameters; $E(z)\equiv H(z)/H_0$; and
 \be{eq14}
 \bar{M}=M+5\log_{10}\left(\frac{cH_0^{-1}}{\rm Mpc}\right)+25
 =M-5\log_{10}h+42.38
 \ee
 is the magnitude zero offset; $h$ is $H_0$ in units of
 100 km/s/Mpc; the absolute magnitude $M$ is assumed to be constant
 after the corrections mentioned above. The data points of the
 latest 182 SNIa Gold dataset compiled in~\cite{r16} are given in
 terms of the distance modulus
 \be{eq15}
 \mu_{obs}(z_i)\equiv m_{obs}(z_i)-M.
 \ee
 On the other hand, the theoretical distance modulus is defined as
 \be{eq16}
 \mu_{th}(z_i)\equiv m_{th}(z_i)-M=5\log_{10}D_L(z_i)+\mu_0,
 \ee
 where
 \be{eq17}
 \mu_0\equiv 42.38-5\log_{10}h.
 \ee
 The theoretical model parameters are determined by minimizing
 \be{eq18}
 \chi^2_{SN}({\bf p})=\sum\limits_{i=1}^{182}
 \frac{\left[\mu_{obs}(z_i)-\mu_{th}(z_i)\right]^2}{\sigma^2(z_i)},
 \ee
 where $\sigma$ is the corresponding $1\sigma$ error; ${\bf p}$
 denotes the model parameters. The parameter $\mu_0$ is a nuisance
 parameter but it is independent of the data points. One can perform
 an uniform marginalization over $\mu_0$. However, there is an
 alternative way. Following~\cite{r17,r18,r19}, the minimization
 with respect to $\mu_0$ can be made by expanding the $\chi^2_{SN}$
 of Eq.~(\ref{eq18}) with respect to $\mu_0$ as
 \be{eq19}
 \chi^2_{SN}({\bf p})=A-2\mu_0 B+\mu_0^2 C,
 \ee
 where
 $$A({\bf p})=\sum\limits_{i=1}^{182}\frac{\left[m_{obs}(z_i)
 -m_{th}(z_i;\mu_0=0,{\bf p})\right]^2}{\sigma_{m_{obs}}^2(z_i)}\,,$$
 $$B({\bf p})=\sum\limits_{i=1}^{182}\frac{m_{obs}(z_i)
 -m_{th}(z_i;\mu_0=0,{\bf p})}{\sigma_{m_{obs}}^2(z_i)}\,,$$
 $$C=\sum\limits_{i=1}^{182}\frac{1}{\sigma_{m_{obs}}^2(z_i)}\,.$$
 Eq.~(\ref{eq19}) has a minimum for $\mu_0=B/C$ at
 \be{eq20}
 \tilde{\chi}^2_{SN}({\bf p})=
 A({\bf p})-\frac{B({\bf p})^2}{C}.
 \ee
 Therefore, we can instead minimize $\tilde{\chi}^2_{SN}$ which is
 independent of $\mu_0$, since
 $\chi^2_{SN,\,min}=\tilde{\chi}^2_{SN,\,min}$ obviously. It is
 worth noting that the corresponding $h$ can be determined by
 $\mu_0=B/C$ for the best-fit parameters.

\begin{center}
 \begin{figure}[htbp]
 \centering
 \includegraphics[width=0.99\textwidth]{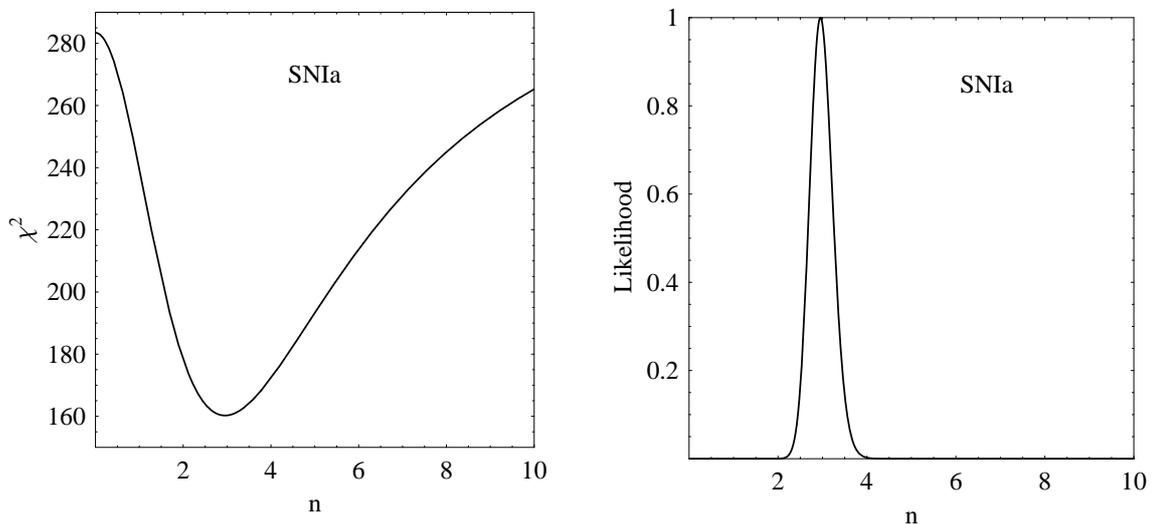}
 \caption{\label{fig2} The $\chi^2=\tilde{\chi}^2_{SN}$
 and the corresponding likelihood for $0<n\leq 10$. These
 results are obtained from the latest 182 SNIa Gold
 dataset compiled in~\cite{r16}.}
 \end{figure}
 \end{center}

In the ADE model~\cite{r11}, there are {\em two} independent model
 parameters, namely, $n$ and $\Omega_{m0}$. To obtain $\Omega_q(z)$
 which is critical in $E(z)$, one should solve the corresponding
 differential equation of $\Omega_q$ with the initial condition
 $\Omega_q(z=0)=\Omega_{q0}=1-\Omega_{m0}$, similar to the case of
 holographic dark energy model (see e.g.~\cite{r23}). In the NADE
 model, however, the situation is different. As mentioned in
 Sec.~\ref{sec2b}, if $n$ is given, we can obtain $\Omega_q(z)$
 from Eq.~(\ref{eq11}) with the initial condition
 $\Omega_q(z_{ini})=n^2(1+z_{ini})^{-2}/4$ at $z_{ini}=2000$ (or
 any $z_{ini}$ which is deep enough into the matter-dominated
 epoch), instead of $\Omega_q(z=0)=\Omega_{q0}=1-\Omega_{m0}$.
 Then, all other physical quantities, such as
 $\Omega_m(z)=1-\Omega_q(z)$ and $w_q(z)$ in Eq.~(\ref{eq10}),
 can be obtained correspondingly. So, $\Omega_{m0}=\Omega_m(z=0)$
 and $\Omega_{q0}=\Omega_q(z=0)$ are {\em not} independent model
 parameters. The only model parameter is $n$. Therefore, the NADE
 model is a {\em single-parameter} model in practice, unlike the
 {\em two-parameters} ADE model~\cite{r11}. To our knowledge, it
 is the third {\em single-parameter} cosmological model besides
 the well-known $\Lambda$CDM model and the DGP braneworld
 model~\cite{r33}. From Eq.~(\ref{eq7}), we have
 \be{eq21}
 E(z)=\left[\frac{\Omega_{m0}(1+z)^3}{1-\Omega_q(z)}\right]^{1/2}.
 \ee
 If the single model parameter $n$ is given, we can obtain
 $\Omega_q(z)$ from Eq.~(\ref{eq11}). And then, we get
 $\Omega_{m0}=1-\Omega_q(z=0)$. Therefore, $E(z)$ is in hand.
 So, we can find the corresponding $\tilde{\chi}^2_{SN}$. In
 Fig.~\ref{fig2}, we present the $\chi^2=\tilde{\chi}^2_{SN}$
 and the corresponding likelihood ${\cal L}\propto e^{-\chi^2/2}$
 for $0<n\leq 10$. We find that the best-fit model parameter is
 $n=2.954$ while $\chi^2_{min}=160.255$. The corresponding
 $h=0.623$. In Table~\ref{tab1}, we present the best-fit value of
 $n$ and the derived $\Omega_{m0}$, $\Omega_{q0}$ and $w_{q0}$
 with $1\sigma$ and $2\sigma$ uncertainties. On the other hand,
 we also fit the $\Lambda$CDM model to the same SNIa dataset and
 find that $\chi^2_{min,\,\Lambda}=158.749$ for the best-fit
 parameter $\Omega_{m0}^\Lambda=0.342$ (the corresponding
 $h=0.626$). So, fitting to the latest 182 SNIa Gold
 dataset~\cite{r16}, the $\Lambda$CDM model is slightly better
 than the NADE model.


 \begin{table}[htbp]
 \begin{center}
 \begin{tabular}{c|c|c|c|c} \hline\hline
 Uncertainty & $n$ & $\Omega_{m0}$ & $\Omega_{q0}$ & $w_{q0}$ \\ \hline
 $1\sigma$ & \ \ $2.954^{+0.264}_{-0.245}$\ \
 & \ \ $0.255^{+0.041}_{-0.038}$\ \
 & \ \ $0.745^{+0.038}_{-0.041}$\ \
 & \ \ $-0.805^{+0.012}_{-0.012}$\ \ \\
 $2\sigma$ & \ \ $2.954^{+0.555}_{-0.475}$\ \
 & \ \ $0.255^{+0.086}_{-0.072}$\ \
 & \ \ $0.745^{+0.072}_{-0.086}$\ \
 & \ \ $-0.805^{+0.023}_{-0.023}$\ \ \\ \hline\hline
 \end{tabular}
 \end{center}
 \caption{\label{tab1} The best-fit value of $n$ and the derived
 $\Omega_{m0}$, $\Omega_{q0}$ and $w_{q0}$ with $1\sigma$ and
 $2\sigma$ uncertainties. These results are obtained from the
 latest 182 SNIa Gold dataset compiled in~\cite{r16}.}
 \end{table}


 \begin{center}
 \begin{figure}[htbp]
 \centering
 \includegraphics[width=0.99\textwidth]{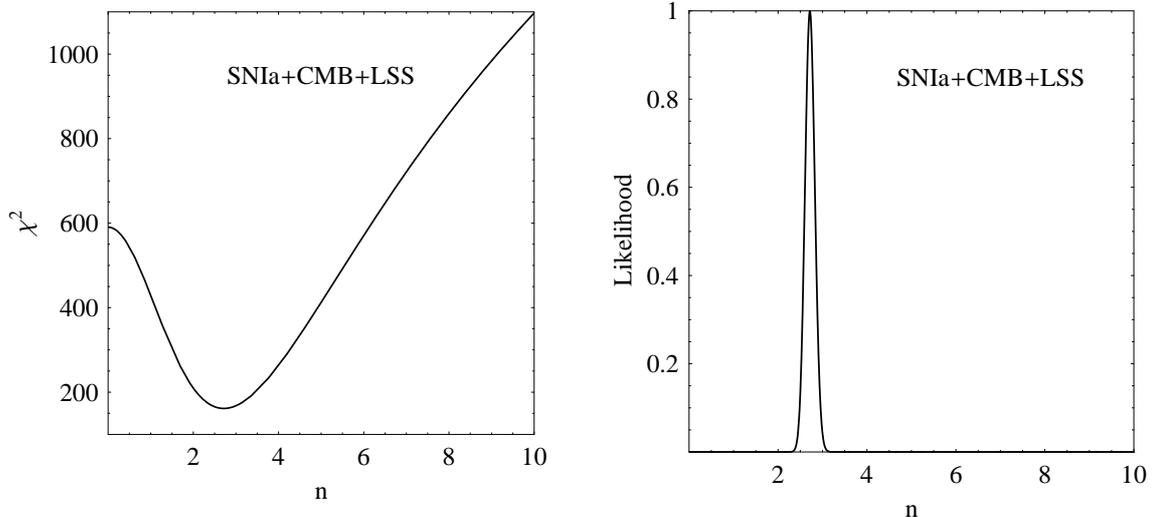}
 \caption{\label{fig3} The $\chi^2$ and the corresponding likelihood
 for $0<n\leq 10$. These results are obtained from the combined SNIa,
 CMB and LSS data.}
 \end{figure}
 \end{center}



\section{Constraints on NADE from SNIa, CMB and LSS}\label{sec4}
Although the constraints on NADE from SNIa alone are fairly tight
 as shown in Table~\ref{tab1}, we can further make the constraints
 tighter by combining SNIa with other complementary observations.
 In the literature, the shift parameter $R$ from CMB~\cite{r20,r21}
 and the parameter $A$ of BAO measurement from LSS~\cite{r22} are
 used extensively, see~\cite{r11,r16,r23} for examples. It is
 commonly believed that both $R$ and $A$ are model-independent and
 contain the essential information of the full CMB and LSS BAO data
 (however, see also e.g.~\cite{r24} and~\cite{r25}). Notice that
 both $R$ and $A$ are independent of Hubble constant $H_0$. In fact,
 the shift parameter $R$ is defined by~\cite{r20,r21}
 \be{eq22}
 R\equiv\Omega_{m0}^{1/2}\int_0^{z_{rec}}
 \frac{d\tilde{z}}{E(\tilde{z})},
 \ee
 where $z_{rec}=1089$ is the redshift of recombination. The shift
 parameter $R$ relates the angular diameter distance to the last
 scattering surface, the comoving size of the sound horizon at
 $z_{rec}$ and the angular scale of the first acoustic peak in CMB
 power spectrum of temperature fluctuations~\cite{r20,r21}. The
 value of $R$ is determined to be $1.70\pm 0.03$~\cite{r21} from
 the WMAP 3-year~(WMAP3) data~\cite{r26}. On the other hand, the
 parameter $A$ of the measurement of BAO peak in the distribution
 of SDSS luminous red galaxies~\cite{r27} is given by~\cite{r22}
 \be{eq23}
 A\equiv\Omega_{m0}^{1/2}E(z_b)^{-1/3}\left[\frac{1}{z_b}
 \int_0^{z_b}\frac{d\tilde{z}}{E(\tilde{z})}\right]^{2/3},
 \ee
 where $z_b=0.35$. In~\cite{r22}, the value of $A$ is determined to
 be $0.469\,(n_s/0.98)^{-0.35}\pm 0.017$, here the scalar spectral
 index $n_s$ is taken to be $0.95$ from the WMAP3 data~\cite{r26}.

We perform a joint analysis of SNIa, the shift parameter $R$ from
 CMB and the BAO peak measurement $A$ from LSS data to constrain
 the NADE model. The total $\chi^2$ is given by
 \be{eq24}
 \chi^2=\tilde{\chi}^2_{SN}+\chi^2_{CMB}+\chi^2_{LSS},
 \ee
 where $\tilde{\chi}^2_{SN}$ is given in Eq.~(\ref{eq20}),
 $\chi^2_{CMB}=(R-R_{obs})^2/\sigma_R^2$ and
 $\chi^2_{LSS}=(A-A_{obs})^2/\sigma_A^2$. In Fig.~\ref{fig3}, we
 present the $\chi^2$ and the corresponding likelihood
 ${\cal L}\propto e^{-\chi^2/2}$ for $0<n\leq 10$. We find that
 the best-fit model parameter is $n=2.716$ while
 $\chi^2_{min}=161.467$. The corresponding $h=0.616$. We present
 the best-fit value of $n$ and the derived $\Omega_{m0}$,
 $\Omega_{q0}$ and $w_{q0}$ with $1\sigma$ and $2\sigma$
 uncertainties in Table~\ref{tab2}. Obviously, these constraints
 in Table~\ref{tab2} are strict enough. On the other hand, we also
 fit the $\Lambda$CDM model to the same combined SNIa, CMB and
 LSS data and find that $\chi^2_{min,\,\Lambda}=162.886$
 for the best-fit parameter $\Omega_{m0}^\Lambda=0.288$ (the
 corresponding $h=0.637$). So, fitting to the combined SNIa, CMB
 and LSS data, the NADE model is slightly better than the
 $\Lambda$CDM model. This makes the NADE model more attractive.

 \begin{table}[tbp]
 \begin{center}
 \begin{tabular}{c|c|c|c|c} \hline\hline
 Uncertainty & $n$ & $\Omega_{m0}$ & $\Omega_{q0}$ & $w_{q0}$ \\ \hline
 $1\sigma$ & \ \ $2.716^{+0.111}_{-0.109}$\ \
 & \ \ $0.295^{+0.020}_{-0.020}$\ \
 & \ \ $0.705^{+0.020}_{-0.020}$\ \
 & \ \ $-0.794^{+0.006}_{-0.005}$\ \ \\
 $2\sigma$ & \ \ $2.716^{+0.224}_{-0.215}$\ \
 & \ \ $0.295^{+0.041}_{-0.038}$\ \
 & \ \ $0.705^{+0.038}_{-0.041}$\ \
 & \ \ $-0.794^{+0.011}_{-0.010}$\ \ \\ \hline\hline
 \end{tabular}
 \end{center}
 \caption{\label{tab2} The best-fit value of $n$ and the derived
 $\Omega_{m0}$, $\Omega_{q0}$ and $w_{q0}$ with $1\sigma$ and
 $2\sigma$ uncertainties. These results are obtained from the
 combined SNIa, CMB and LSS data.}
 \end{table}


\section{Conclusion}\label{sec5}
In this work, we consider the NADE model proposed in~\cite{r12},
 which might have some physical motivations connected to the
 quantum fluctuations of spacetime. Thanks to its special analytic
 features in the radiation-dominated and matter-dominated epochs,
 NADE is a {\em single-parameter} model in practice, unlike the
 {\em two-parameters} ADE model~\cite{r11}. If the single model
 parameter $n$ is given, all other physical quantities of NADE can
 be determined correspondingly. To our knowledge, it is the third
 {\em single-parameter} cosmological model besides the well-known
 $\Lambda$CDM model and the DGP braneworld model~\cite{r33}. We
 find that the coincidence problem could be solved naturally in
 the NADE model provided that the single model parameter $n$ is
 of order unity. In addition, we constrain NADE by using the
 cosmological observations of SNIa, CMB and LSS. The joint
 analysis gives the best-fit parameter (with $1\sigma$ uncertainty)
 $n=2.716^{+0.111}_{-0.109}$. The derived $\Omega_{m0}$,
 $\Omega_{q0}$ and $w_{q0}$ (with $1\sigma$ uncertainties) are
 $0.295^{+0.020}_{-0.020}$, $0.705^{+0.020}_{-0.020}$ and
 $-0.794^{+0.006}_{-0.005}$, respectively. These constraints are
 strict enough. On the other hand, fitting to the combined SNIa,
 CMB and LSS data, we find that the NADE model is slightly better
 than the $\Lambda$CDM model. This makes the NADE model more
 attractive.

If we add other observations such as Chandra X-ray
 observations~\cite{r28}, observational $H(z)$ data~\cite{r29},
 the lookback time data compiled in~\cite{r30} and so on, it is
 expected that the constraints on NADE model will be very tight
 and the resulting parameters can be used to make some exact
 predictions of NADE. Of course, the other SNIa datasets such as
 SNLS~\cite{r31} and ESSENCE~\cite{r32} are also useful to constrain
 the NADE model. It is worth noting that in this work we only used
 the parameters $R$ and $A$ from WMAP3 and SDSS data, the resulting
 constraint on $n$ is primary in some sense, and therefore it is
 of great interest to constrain the NADE model by using the global
 fitting to the full CMB and LSS data via Markov Chain Monte
 Carlo (MCMC) analysis. This is an issue which deserves further
 investigation in the future.


\section*{ACKNOWLEDGMENTS}
We thank the anonymous referees for quite useful comments and
 suggestions, which help us to improve this work. We are grateful
 to Prof. Shuang~Nan~Zhang for helpful discussions. We also
 thank Minzi~Feng, as well as Hui~Li, Yi~Zhang, Xing~Wu,
 Li-Ming~Cao, Xin~Zhang, Jian~Wang and Bin~Hu, for kind help
 and discussions. This work was supported in part by a grant from
 China Postdoctoral Science Foundation, a grant from Chinese
 Academy of Sciences~(No.~KJCX3-SYW-N2), and by NSFC under grants
 No.~10325525, No.~10525060 and No.~90403029.

\newpage 

\renewcommand{\baselinestretch}{1.2}


\end{document}